\documentclass[aps,prb,twocolumn,showpacs,tightenlines]{revtex4}
\usepackage{epsfig}
\usepackage{graphicx}
\usepackage{amsmath}
\usepackage{amsfonts}
\usepackage{amssymb}
\usepackage{bm}
\begin{document}
\title{Quantum Criticality in Dimerized Spin Ladders}
\author{Gennady Y. Chitov}
\affiliation{Department of Physics, Laurentian University, Ramsey
Lake Road, Sudbury, ON, P3E 2C6, Canada }
\author{Brandon W. Ramakko}
\affiliation{Department of Physics, Laurentian University, Ramsey Lake Road,
Sudbury, ON, P3E 2C6, Canada }
\author{Mohamed Azzouz}
\altaffiliation{Also at School of Science and Engineering, Al Akhawayn
University, Ifrane 53000, Morocco.}
\affiliation{Department of Physics,
Laurentian University, Ramsey Lake Road, Sudbury, ON, P3E 2C6, Canada }
\date{\today}

\begin{abstract}
We analyze a possibility of quantum criticality (gaplessness) in dimerized
antiferromagnetic two- and three-leg spin-$\frac12$ ladders. Contrary to
earlier studies of these models, we examine different dimerization patterns in
the ladder. We find that ladders with the columnar dimerization order have
lower zero-temperature energies and they are always gapped. For the staggered
dimerization order, we find the quantum critical lines, in agreement with
earlier analyses. The bond mean-field theory we apply, demonstrates its
quantitative accuracy and agrees with available numerical results. We conclude
that unless some mechanism for locking dimerization into the energetically less
favorable staggered configuration is provided, the dimerized ladders do not
order into the phase where the quantum criticality occurs.
\end{abstract}

\pacs{ 75.10.Jm, 75.10.-b,  75.10.Pq, 73.43.Nq, 64.60.-i}

\maketitle
%
%
\section{Introduction}\label{Intro}
%
%
There has been a lot of interest in spin ladders for more than a decade by now,
mainly due to their very intriguing critical properties. One of the most
peculiar ones is that the existence of a gap (i.e. mass) depends on the number
of legs. The spin excitations in a $m$-leg spin ladder are gapped if $m$ is
even, and the system is gapless (quantum critical) when the number of legs $m$
is odd. For reviews see Ref.[\onlinecite{Dagotto96,Giamarchi04}]. The
even-$m$-leg ladders provide a very interesting example of systems where the
gap (mass) generation is not accompanied by a long-range order or apparent
symmetry breaking. Such systems, known as spin liquids, are under enormous
scrutiny. They are notoriously difficult to realize in dimension more than one,
and are believed to be relevant to the physics of high-$T_c$ superconductors.
\cite{Lee06} Broadly speaking, spin ladders are interesting for studying gap
(mass) generation: needless to say that this a very deep question in all
physics.

The particular question we address in this study is a possibility of ``restored
quantum criticality" (or gaplessness) in a system (ladder) built from gapped
blocks (dimerized chains).
\cite{Delgado96,Delgado98,Kotov99,Cabra99,Nersesyan00,Okamoto03,Almeida07a,Almeida07b}
Let us explain the issue taking a dimerized two-leg ladder as an example.
\cite{noteAF} It is well known that a single Heisenberg spin-$\frac12$ chain
with alternating spin exchange (a.k.a. the dimerized Heisenberg chain) is
gapped. \cite{Giamarchi04} When two chains are coupled into a ladder, the
system is gapped even without dimerization. However, quite remarkably and
counterintuitively, the dimerized two-leg ladder can be gapless, as was
conjectured in Ref.[\onlinecite{Delgado96}]. Subsequent numerical work
\cite{Delgado98,Kotov99,Cabra99,Okamoto03} confirmed the critical (gapless)
line proposed by Martin-Delgado \textit{et al}. \cite{Delgado96} Similarly, the
originally conjectured critical line in a dimerized three-leg ladder
\cite{Delgado96} was confirmed in the very recent DMRG calculations.
\cite{Almeida07a,Almeida07b}

We need to stress one very important point: in the earlier work
\cite{Delgado98,Kotov99,Cabra99,Nersesyan00,Okamoto03,Almeida07a,Almeida07b} on
the \textit{intrinsically dimerized} ladders, the intra-chain dimerization as
well as the (staggered) dimerization ordering pattern of the whole ladder (cf.
Figs. \ref{Dim_Lad},\ref{3LDim_Lad} below) are taken for granted, i.e., as the
\textit{model built-in assumptions}. In the absence of a physical mechanism
which would lock a ladder with dimerized chains into a particular dimerization
order, it is natural and physically more reasonable to consider various
possible ordering patterns on the same footing. It means that the possibility
of quantum criticality (gaplessness) has to be addressed in this broader
setting. A strong motivation comes from a recent work on the three-leg ladder
coupled to phonons.\cite{Khalada} It turns out that the phonon-induced
intrachain dimerizaton,  appearing in this ladder at the spin-Peierls
transition, occurs into the columnar dimerization order (cf. Fig.
\ref{3LDim_Lad} below). The latter, and not the staggered order, minimizes the
spin-phonon Hamiltonian. The columnar dimerized three-leg ladder is always
gapped.

So, we begin from the study of the dimerized two-leg ladder in Sec.~\ref{2leg}.
From analyses of the limits where the behavior of this ladder is known exactly,
we conjecture that the gapped columnar dimerized phase has lower energy. This
is confirmed by the subsequent treatment within the bond mean-field theory.
\cite{Azzouz93,Azzouz94,Kofi} The latter approach, when applied for the case of
staggered dimerization, demonstrates a good agreement with available numerical
results, and also yields quite simple formulas for analytical treatments of the
problem. A similar program with analogous conclusions is carried out for the
three-leg ladder in Sec.~\ref{3leg}.
%
%
\section{Two-Leg Ladder}\label{2leg}
\begin{figure}[h]
\epsfig{file=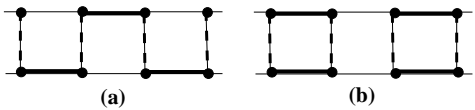,width=0.35\textwidth,angle=0} \caption{Dimerized
two-leg ladder. Bold/thin/dashed lines represent the stronger/weaker chain
coupling $J(1 \pm \delta)$ and rung coupling $J_\bot$, respectively.
Dimerization patterns: (a) - staggered; (b)- columnar.} \label{Dim_Lad}
\end{figure}

%
%
We consider a two-leg dimerized spin ladder given by the Hamiltonian:
\begin{eqnarray}
\label{Ham}
 H_{\mathrm{2L}} &=& \sum_{\alpha=1,2} \sum_{n=1}^N  J_{\alpha} (n)
 \mathbf{S}_{\alpha} (n) \cdot  \mathbf{S}_{\alpha} (n+1) \cr
 &+& J_\bot \sum_{n=1}^N \mathbf{S}_1(n) \cdot \mathbf{S}_2(n).
\end{eqnarray}
The $m$-leg ladder has $N$ rungs and $m \cdot N$ spins. We consider the
situation when dimerization occurs along the chains ($\alpha=1,2$) only, while
the coupling $J_\bot$ is constant on each rung. One needs to consider two
possible dimerization patterns: alternated, when staggering occurs in both
directions
\begin{equation}
\label{Jalt}
 J_{\alpha} (n)=J[1+(-1)^{n+\alpha}\delta],
\end{equation}
and columnar, when
\begin{equation}
\label{Jcol}
 J_{\alpha} (n)=J[1+(-1)^{n}\delta].
\end{equation}
The two patterns are shown in Fig.\ \ref{Dim_Lad}
%
%
\subsection{Analytical Limits}\label{AnLim}
%
%
Let us discuss the points on the $(\delta,J_\bot)$-plane where the properties
of the model (\ref{Ham}) are known.

$\delta=0,~J_\bot=0$ --  two decoupled critical (gapless) Heisenberg
chains.

$\delta=0,~J_\bot>0$ -- uniform gapped ladder.

\begin{figure}[h]
\epsfig{file=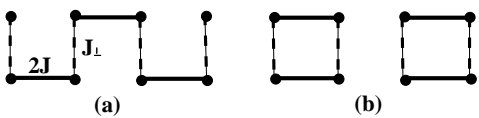,width=0.35\textwidth,angle=0} \caption{Completely
dimerized ladder, $\delta =1$. (a): Alternated staggering reduces the model
(\ref{Ham}) to a snake-like dimerized Heisenberg chain of $2N$ spins; (b):
Columnar order degenerates into a set of $N/2$ decoupled plaquettes.}
\label{Tot_Dim}
\end{figure}

$\delta=1$ -- complete dimerization. In this limit the model reduces
to: \\
\textit{(a)} the snake-like dimerized Heisenberg chain of $2N$ spins
in case of the alternated staggered pattern (Fig.\ \ref{Tot_Dim}a)\\
\textit{(b)} a set of $N/2$ decoupled four-spin plaquettes in case
of the columnar dimerization pattern (Fig.\ \ref{Tot_Dim}b)
For a four-spin plaquette where its nearest-neighbor spins interact
via the Heisenberg couplings $J_1$ and $J_2$, the spectrum is
known.\cite{Bose03} The ground-state singlet has the energy
\begin{equation}
\label{eplaq}
\mathcal{E}_{\text{\tiny
$\square$}}=-\frac{J_1+J_2}{2}-\sqrt{J_1^2+J_2^2-J_1J_2},
\end{equation}
and is separated by a gap from the closest triplet state with with
the energy $\mathcal{E}_t=-\frac{J_1+J_2}{2}$.

As was first noted in Ref.~\onlinecite{Delgado96}, at the critical point
$2J=J_\bot \equiv J_c$ the snake-like chain (Fig.\ \ref{Tot_Dim}a) becomes just
a uniform (\textit{gapless}) Heisenberg model. Its  ground-state energy is
\begin{equation}
\label{Ehc} E_{\mathrm{HC}}=-2\Big( \log 2-\frac14 \Big) J_c N \approx -0.88J_c
N.
\end{equation}
The set of $N/2$ plaquettes (\ref{eplaq}) has a lower ground-state
energy
\begin{equation}
\label{Eplaq} E_{\text{\tiny $\square$}}=-J_c N.
\end{equation}
The simple results (\ref{Ehc},\ref{Eplaq}) question the very existence of the
critical line predicted in \cite{Delgado96}. That line was conjectured
essentially from a continuity argument and analysis of the model in the two
solvable (integrable) points $\delta=0,~J_\bot=0$ and $\delta=1,~J_\bot= 2 J$.
The latter was assumed to be a critical (gapless) point, since the alternated
staggering pattern (thus the snake-like chain in the limit $\delta =1$) was
assumed to be a ground state. The subsequent analyses addressed the issue of
the critical line (within the alternated staggering pattern) more
quantitatively, \cite{Delgado98,Kotov99,Cabra99,Nersesyan00,Okamoto03} while,
to the best of our knowledge, comparisons with the energy of the columnar
pattern were not done. As one can see from (\ref{Ehc},\ref{Eplaq}), the gapped
plaquettes have lower energy than the critical (gapless) snake-like Heisenberg
chain. Note that the decoupled plaquettes at $\delta =1$ evolve smoothly from
the columnar dimerization pattern at $\delta < 1$.

Let us now look at the line $\delta=1$, $J_\bot \neq 2 J$. It is
convenient to parameterize couplings as
\begin{eqnarray}
\label{Jpl}
2J &\equiv& J_\circ (1+\delta_{\mathrm{eff}}), \\
\label{Jm} J_\bot &\equiv& J_\circ (1-\delta_{\mathrm{eff}}).
\end{eqnarray}
The plaquette pattern results in the ground-state energy
\begin{equation}
\label{EplNU} E_{\text{\tiny $\square$}}=- \frac12 J_\circ N \Big( 1+\sqrt{1+3
\delta_{\mathrm{eff}}^2 } \Big).
\end{equation}
The energy minimum of the snake-like configuration is given by the ground-state
energy of the dimerized Heisenberg chain. The latter is known to be equivalent
to the massive sine-Gordon (integrable) model, perturbed by the marginal
(logarithmic) corrections. \cite{BE81,Affleck89} A leading order ansatz reads
as
\begin{equation}
\label{Edhc} E_{\mathrm{DHC}} \approx -2 J_\circ N  \Big( \log 2-\frac14 +
a_\circ \frac{\delta_{\mathrm{eff}}^{4/3}}{ \ln
\frac{\delta_{\circ}}{\delta_{\mathrm{eff}} } } \Big)~,
\end{equation}
where from the recent numerical calculations \cite{Papen03} $a_\circ \approx
2.2$ and $\delta_{\circ} \approx 110$. To get a more accurate analytical
expression for $E_{\mathrm{DHC}}$ is an involved problem (for a recent analysis
and more references, see \cite{Orignac}). However numerically, the ansatz
(\ref{Edhc}) or even the unperturbed sine-Gordon formula \cite{Orignac}
\begin{equation}
\label{ESG} E_{\mathrm{SG}} \approx -2 J_\circ N  \Big( \log 2-\frac14 + 0.2728
\delta_{\mathrm{eff}}^{4/3} \Big)
\end{equation}
both work quite well. A direct inspection of
Eqs.(\ref{EplNU},\ref{Edhc},\ref{ESG}) (e.g., plots) clearly shows that the
plaquettes again provide the lower energy state on the line $\delta=1$ near the
point $J_\bot = 2 J \Longleftrightarrow \delta_{\mathrm{eff}}=0$.\cite{noteDH}
%
%
\subsection{Mean-Field Equations}\label{MFE}
%
%
Away from integrable points we need to resort to approximations. We will treat
the dimerized ladders in the framework of the bond mean-field
theory.\cite{Azzouz93, Azzouz94} This approach for the case of dimerized
three-leg ladder is described in detail in \cite{Khalada}, and the case of two
legs is essentially a simplified version of the former. First, the spin
Hamiltonian is mapped onto an interacting fermion problem via a 2D version of
the Jordan-Wigner transformation, proposed by one us. \cite{Azzouz93} Then, the
phase differences due to hopping of the fermions around any given elementary
plaquette is approximated by $\pi$. The quartic fermionic terms $ c_{i,}^\dag
c_{i,\alpha}c_{i+1,\alpha}^\dag c_{i+1,\alpha}$ are treated within the
Hartree-Fock approximation, i.e., are decoupled using the single-particle
(bond) parameters. The latter are defined as
\begin{equation}
\label{Q}
    Q_+ =\langle
 c_{2i,\alpha}c_{2i+1,\alpha}^\dag\rangle ,~Q_- =\langle
 c_{2i+1,\alpha}c_{2i+2,\alpha}^\dag\rangle,
\end{equation}
with $\alpha=1,2$. A single bond parameter suffices in the rung direction:
$P=\langle c_{2i,\alpha}c_{2i,\alpha+1}^\dag\rangle$. Fourier transforming
along the chain direction and using the Nambu formalism, the mean-field theory
results in the single-particle effective Hamiltonian
\begin{equation}
\label{H}
  H^{\rm s/c} = \sum_k\Psi_k^\dag\mathcal{ H}^{(\rm s/c)} \Psi_k+N C_2,
\end{equation}
where the Hamiltonian density $\mathcal{H}^{\rm s/c}$ is a $ 4\times 4$ matrix
and the Nambu spinor ${\Psi_k^\dag} \equiv
 \left(\begin{array}{cccccc} c_{1k}^{A\dag}&c_{1k}^{B\dag}& c_{2k}^{A\dag}&
 c_{2k}^{B\dag}
 \end{array}\right).$
Here $c_{\alpha k}^{\sharp}$ is the Fourier transform of $c_{i
\alpha}^{\sharp}$. To account for the dimerization (doubling of the lattice
spacing), the lattice is subdivided into two sublattices A and B. The explicit
form of the effective Hamiltonian (\ref{H}) depends on the dimerization pattern
(staggered or columnar) which is accounted for by an extra label (s,c) in the
above equation, and
\begin{equation}
\label{C2}
    C_2 = J_+| Q_+|^2+ J_- |Q_-| ^2 + J_\bot  |P|^2~.
\end{equation}
Diagonalization of $\mathcal{H}^{\rm s/c}$ yields four energy eigenvalues $\pm
E^{\rm s}_n(k)$, $n=1,2$:
\begin{equation}
\label{Es122L}
 E^{\rm s}_{1/2}(k)=\frac12 \sqrt{
  W+ J_{\bot 1}^2 \mp  2 J_{\bot 1} (J_{1+}-J_{1-})\cos k
 } ~,
\end{equation}
for the staggered phase. We define
\begin{eqnarray}
 \label{J1s}
 J_{1\pm} &=& J(1\pm\delta)(1+2Q_\pm), ~
 J_{\bot 1}= J_\bot(1+2P),~~~\\
 \label{W}
 W &=& J_{1+}^2+ J_{1-}^2 - 2 J_{1+} J_{1-} \cos(2k).
\end{eqnarray}
For the columnar pattern we obtain two doubly degenerate energy eigenvalues $
\pm E^{\rm c}(k)$:
\begin{equation}
\label{Ec2L}
 E^{\rm c}(k)=\frac12 \sqrt{W+ J_{\bot 1}^2 } ~,
\end{equation}
The partition function of the single-particle Hamiltonian (\ref{H}) can be
found in a closed form, and the free energy per spin is
\begin{equation}
\label{FrEn}
 F^{\sharp}= \frac1m C_m -\frac{\log 2}{\beta}-
\frac{2 \mathfrak{g}}{m \pi \beta}
 \sum_n \int_0^{\frac{\pi}{2}}  \log \cosh \frac{\beta E^{\sharp}_n (k)}{2} d k ~,
\end{equation}
where $\beta=1/ {k_B T}$ and $\mathfrak{g}$ is eigenvalue's degeneracy. We gave
the above formula valid for $m$ legs. The mean-field equations are derived from
minimization of the free energy (\ref{FrEn}) with respect to the mean-field
parameters $Q_\pm$ and $P$. These self-consistent integral equations are solved
numerically.
%
%
\subsection{Analysis}\label{Analyze}
\begin{figure}[h]
\epsfig{file=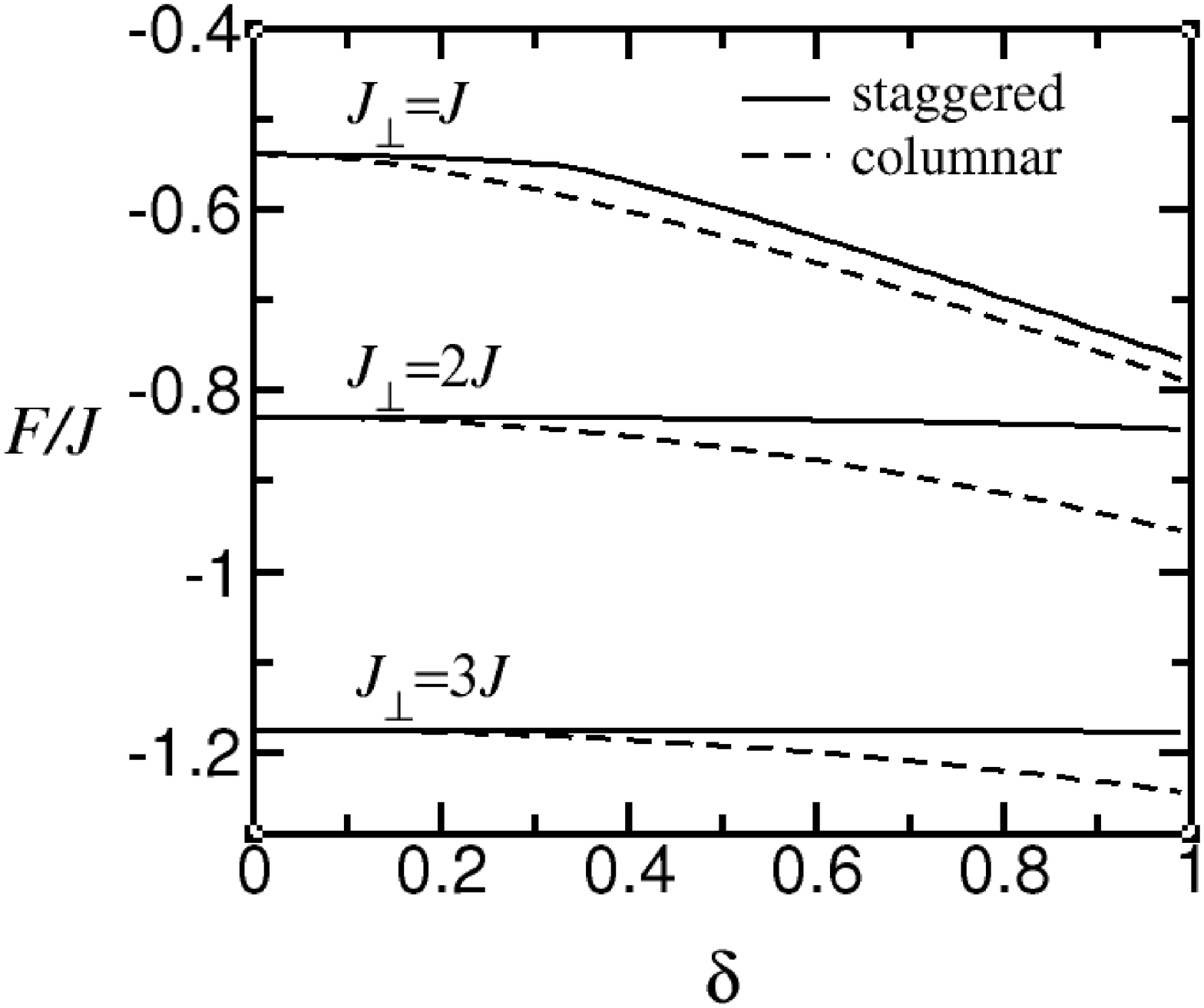,width=0.35\textwidth,angle=0} \caption{Two-leg
ladder: Energies of staggered and columnar dimerization configurations at
$T=0$.} \label{FEn2L}
\end{figure}

%
%
To start with, let us make the following observation: In the antiferomagnetic
two-leg ladder  (\ref{Ham}) with intrinsically dimerized chains we cannot think
of a particular mechanism to enforce a specific (i.e., staggered or columnar)
dimerization pattern. So, having the dimerized chains, the ladder must admit
the pattern which minimizes its free energy. Using Eq.~(\ref{FrEn}), we
compared the zero-temperature energies of two configurations in various
parameter ranges. The characteristic results are given in Fig.\ \ref{FEn2L}.
The columnar pattern corresponds the lower energies in all cases, so it is the
thermodynamically stable state (phase).
Now let us analyze the excitation spectra of two phases. We will
identify the minimal band gap of the eigenvalues (\ref{Es122L}) or
(\ref{Ec2L}) with the spin gap of the corresponding phase. To
demonstrate that this is true, it suffices, e.g., to calculate the
uniform spin susceptibility from the free energy (\ref{FrEn}) with
an external magnetic field added.

We find that the columnar phase is \textit{always gapped } with the
gap $\Delta_c= E^{\rm c}(0)$ given by
\begin{equation}
\label{Dc2}
    \Delta_c= J \sqrt{ u^2 \delta^2 +  \Big(\frac{J_{\bot} }{2 J} \Big)^2 p^2}
    ~,
\end{equation}
where
\begin{eqnarray}
 \label{s}
 u &=& 1+(Q_+ + Q_-) + \frac{Q_+ - Q_-}{\delta},~~~\\
 \label{p}
 p &=& 1+2P.
\end{eqnarray}
Qualitatively, the columnar phase is similar to the uniform two-leg
ladder, i.e. gapped. The gap (\ref{Dc2}) persists in the limit of
decoupled chains $J_\bot \to 0$, as it must be, and disappears only
together with the vanishing chain dimerization $\delta \to 0$. If
the latter limit is taken first, then the uniform ladder is gapped
with $\Delta_c=J_\bot p/2$.

\begin{figure}
\epsfig{file=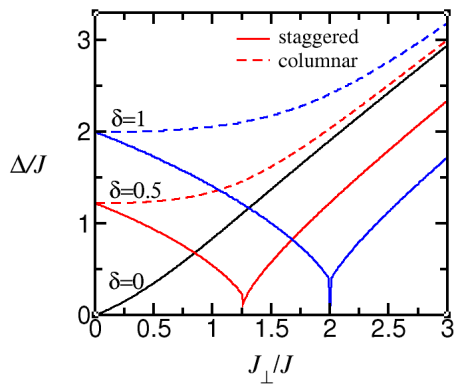,width=0.35\textwidth,angle=0} \caption{Two-leg
ladder: Gaps of the staggered and columnar phases.} \label{Gaps2L}
\end{figure}

The staggered phase is more interesting. Its gap $\Delta_s= E^{\rm s}_1(0)$
reads as: \cite{noteG}
\begin{equation}
\label{Ds2}
    \Delta_s= J \big| u \delta -  \frac{J_\bot }{2 J} p \big|
    ~.
\end{equation}
\textit{It vanishes on a certain critical line, even if neither
dimeritation $\delta$, nor rung coupling $J_\bot$ are zero.} Thus,
the mean-field theory confirms the earlier conjecture
\cite{Delgado96}, which has been so far corroborated only by
numerical calculations. \cite{Delgado98,Kotov99,Cabra99,Okamoto03}
With the bond parameters $Q_\pm, P$ determined from the mean-field
equations, we plot the gaps $\Delta_{c,s}$ given by
Eqs.~(\ref{Dc2},\ref{Ds2}) as functions of coupling ratio $J_\bot
/J$ for  several dimerizations in Fig.\ \ref{Gaps2L}.
Note that the gaps $\Delta_c$ and $\Delta_s$ coincide at``the decoupled chains
line" $J_\bot /J=0$, as well as in the absence of dimerization. The qualitative
differences between the gaps in the columnar and staggered phases are clearly
seen. The former is always non-zero if the chains are dimerized, and increases
with the growth of $J_\bot /J$. The latter behaves non-monotonously, passing
though a critical point. It is worth noting that the known analytical limits
(discussed above) are recovered exactly by the mean-field equations. In
particular, on the ``decoupled" axis $J_\bot =0$, the mean-field parameter
$Q_\pm$ whose physical meaning is the average of the dimerization operator
$\mathbf{S}_{\alpha} (n) \cdot  \mathbf{S}_{\alpha} (n+1)$, behaves almost as
that in the dimerized $XY$-chain, coinciding with the exact values of that
quantity at the ends of the interval $\delta \in [0,1]$. For the dimerized
$XY$-chain
\begin{equation}
\label{QpmXY}
    Q_\pm= \langle \mathbf{S}_{n} \cdot  \mathbf{S}_{n+1} \rangle \equiv
    t\pm \delta \cdot \eta
\end{equation}
the uniform term $t$ and the dimerization susceptibility $\eta$ at $T=0$ are
given in terms of the elliptic functions and can be found, e.g., in
Ref.~\onlinecite{ChitovPRB04}. In particular,
\begin{eqnarray}
 \label{Qd0}
 \delta &=& 0~: ~~Q_\pm  = \frac{1}{\pi},~~~\\
 \label{Qd1}
 \delta &=& 1~: ~~Q_+= \frac12~,~~Q_-=0.
\end{eqnarray}
For instance, at $\delta =1$ ($J_\bot =0$), the ladder reduces to the set of
decoupled dimers with the Heisenberg exchange $2J$. Our
Eqs.~(\ref{Dc2}-\ref{Qd1}) result in the exact value of the gap $\Delta=2 J$
for this case. Similar, at the integrable quantum critical point $\delta =1$,
$J_\bot =2J $ (uniform snake-like chain), we have $Q_+ =P$, and the mean-field
(\ref{Ds2}) correctly predicts a vanishing gap.

\begin{figure}
\epsfig{file=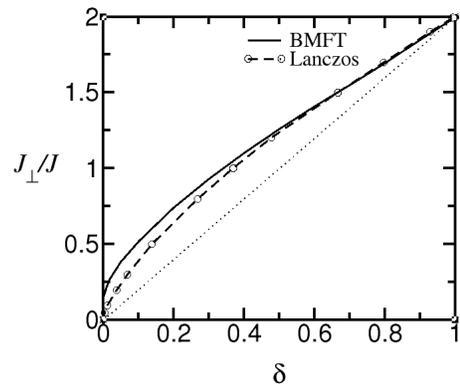,width=0.35\textwidth,angle=0} \caption{Two-leg
ladder: Critical line where the gap of the staggered phase vanishes. Along with
the present bond mean-field theory predictions, the numerical diagonalization
data from Ref.~\onlinecite{Okamoto03} are shown.} \label{Crit2L}
\end{figure}

We plot in Fig.\ \ref{Crit2L} the mean-field critical line given by
\begin{equation}
\label{Cr2L}
     \frac{J_\bot }{J}=  \frac{2 u}{p} \delta~,
\end{equation}
along with the numerical results of Ref.~\onlinecite{Okamoto03}.
As one can see, the mean field works quite well quantitatively. It is
surprising, but the mean field does much better than, e.g., the non-linear
sigma model (NL$\sigma$M). The latter fails to predict the correct critical
line.\cite{Delgado96} Analysis of the topological term of the NL$\sigma$M
predicts that the line ends at $\delta = \frac12$, $J_\bot=0$. The dotted line
in Fig.\ \ref{Crit2L} corresponds to the case when in the gap equations we set
the bond parameters to their maximal values $Q_\pm=P=\frac12$ (i.e., $u=p=2$),
so $J_\bot /J =2 \delta$. Such a naive ``uniform isotropic limit'' simplifies
drastically the formulas, and provides a decent approximation.
%
%
\section{Three-Leg Ladder}\label{3leg}
\begin{figure}[h]
\epsfig{file=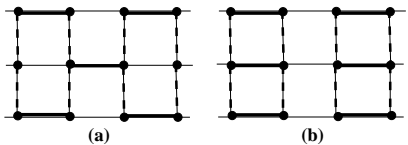,width=0.30\textwidth,angle=0} \caption{Dimerized
three-leg ladder. Line notations are the same as in Fig.\ \ref{Dim_Lad}.
Dimerization patterns: (a) - staggered; (b)- columnar. } \label{3LDim_Lad}
\end{figure}

%
%
A three-leg dimerized ladder is defined by the Hamiltonian:
\begin{eqnarray}
\label{Ham3}
 H_{\mathrm{3L}} &=& \sum_{\alpha=1,2} \sum_{n=1}^N  J_{\alpha} (n)
 \mathbf{S}_{\alpha} (n) \cdot  \mathbf{S}_{\alpha} (n+1) \cr
 &+& J_\bot \sum_{n=1}^N
 \big[ \mathbf{S}_1(n)\cdot  \mathbf{S}_2(n)+\mathbf{S}_2(n) \cdot \mathbf{S}_3(n) \big].
\end{eqnarray}
Similar to the two-leg case, we consider dimerizations of the whole ladder
(\ref{Jalt},\ref{Jcol}) which follow the staggered or columnar patterns, as
shown in Fig.\ \ref{3LDim_Lad}.

%
Contrary to the previous case, now we don't have limits where the behavior of
this ladder is known \textit{exactly}. Although the non-dimerized limit $\delta
=0$ is known to be gapless, and the three-leg ladder, roughly speaking, can be
described as a ``renormalized" spin-$\frac12$ Heisenberg chain.
\cite{Dagotto96,Kofi} One can easily see from Fig.\ \ref{3LDim_Lad} that in the
totally dimerized limit $\delta =1$ the columnar dimerized ladder becomes a set
of $N/2$ decoupled 6-spin plaquettes, while the staggered pattern evolves into
a ``decorated" chain. The latter, to the best of our knowledge, has been not
analyzed earlier.

We treat the problem in the bond mean-field approach as above. In fact, we will
use directly our earlier results for the three-leg ladder coupled to phonons
\cite{Khalada}, with minor modifications.  Along with one couple of the bond
parameters $Q_\pm$, defined by (\ref{Q}) with $\alpha =1,3$, one needs an extra
couple of parameters $Q_\pm^\prime$ for the chain in the middle, defined by the
same equation with $\alpha =2$. The effective Hamiltonian density
$\mathcal{H}^{\rm s/c}$ (cf. Eq.~(\ref{H})) is now a $ 6\times 6$ matrix, and
the Nambu spinor ${\Psi_k^\dag}=
 \left(\begin{array}{cccccc} c_{1k}^{A\dag}&c_{1k}^{B\dag}& c_{2k}^{A\dag}&
 c_{2k}^{B\dag}& c_{3k}^{A\dag}& c_{3k}^{B\dag}
 \end{array}\right).$ The constant term
\begin{eqnarray}
\label{C3}
    C_3 &=& J_+| Q_+|^2+ J_- |Q_-| ^2+ \frac12 J_- |Q'_-|^2 \cr
   &+& \frac12 J_+ |Q'_+|^2+2J_\bot  |P|^2~.
\end{eqnarray}
Diagonalization of $\mathcal{H}^{\rm s/c}$ results in six energy eigenvalues $
\pm E^{\rm s/c}_j(k)$, $j=1,2,3$ for each configuration:
\begin{equation}
\label{E1}
    E^{\rm s}_1(k) =E^{\rm c}_1(k) = \frac12 \sqrt {W},
\end{equation}
and
\begin{widetext}
\begin{equation}
\label{E23}
 E^{\rm s/c}_{n}(k)= \frac{1}{2^{\frac32} } \left[(-1)^n\sqrt
 {(W-W')^2+8J_{\bot 1}^2(W+W'+2Y^{\rm s/c}) }+W+W'+4J_{\bot 1}^2
 \right]^\frac12, ~~n=2,3,
\end{equation}
\end{widetext}
where
\begin{eqnarray}
\label{tsc}
 Y^{\rm s}&=& \big(J_{1+}J'_{1+}+J_{1-}J'_{1-}\big)\cos
(2k)-\big(J_{1+}J'_{1-}+J_{1-}J'_{1+}\big), \cr
 Y^{\rm c}&=& \big(J_{1+}J'_{1-}+J_{1-}J'_{1+}\big)\cos
(2k)-\big(J_{1+}J'_{1+}+J_{1-}J'_{1-}\big), \nonumber
\end{eqnarray}
and $J'_{1 \pm}$ and $W'$ are given by Eqs.(\ref{J1s},\ref{W}) where $Q_\pm \to
Q_\pm^\prime$. The mean-field equations are obtained via minimization of the
free energy, given by Eq.~(\ref{FrEn}) with $m=3$. Our analysis here follows
the steps of the above two-leg case, so we will be brief.

\begin{figure}[h]
\epsfig{file=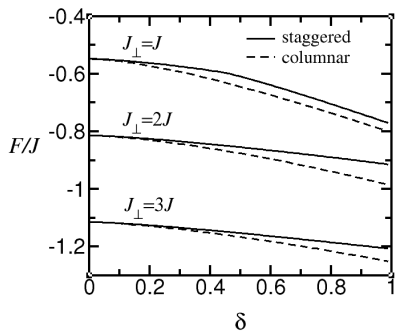,width=0.30\textwidth,angle=0} \caption{Three-leg
ladder: Energies of staggered and columnar dimerization configurations at
$T=0$.} \label{FEn3L}
\end{figure}

Similar to the two-leg case, the columnar order corresponds to a state with
lower energy of the dimerized three-leg ladder, thus this is a
thermodynamically stable state. We present several plots of the
zero-temperature energies for the two types of dimerization order in
Fig.~\ref{FEn3L}.
%
%
The columnar phase of the three-leg ladder is \textit{always gapped}
when $\delta \neq 0$. The gap $\Delta_c= E^{\rm c}_1(0)$ is given by
\begin{equation}
\label{Dc3}
    \Delta_c= J u  \delta ~,
\end{equation}
In the limit $\delta \to 0$, this mean-field approach correctly predicts
gaplessness \cite{Kofi}, in accordance with the general results for odd-$m$-leg
ladders. \cite{Dagotto96,Giamarchi04} We should however point out that the
gapless limit $\Delta_c \to 0$ when $\delta \to 0$ is recovered only if we set
$(Q^\sharp_-  - Q^\sharp_+) \to 0$. The latter limit, obviously correct from
physical point of view [cf. Eq.(\ref{QpmXY})], is not automatically fulfilled
in the bond mean-field equations. This is an artefact of the particular bond
mean-field decoupling scheme, and is not intrinsic for any mean-field
decoupling, as one see, e.g., from the (classical) mean-field results of
Ref.~\onlinecite{Bulaev63}. It is worth noting that the simple analytical
approximations we derive, are free of these flaws.

\begin{figure}[h]
\epsfig{file=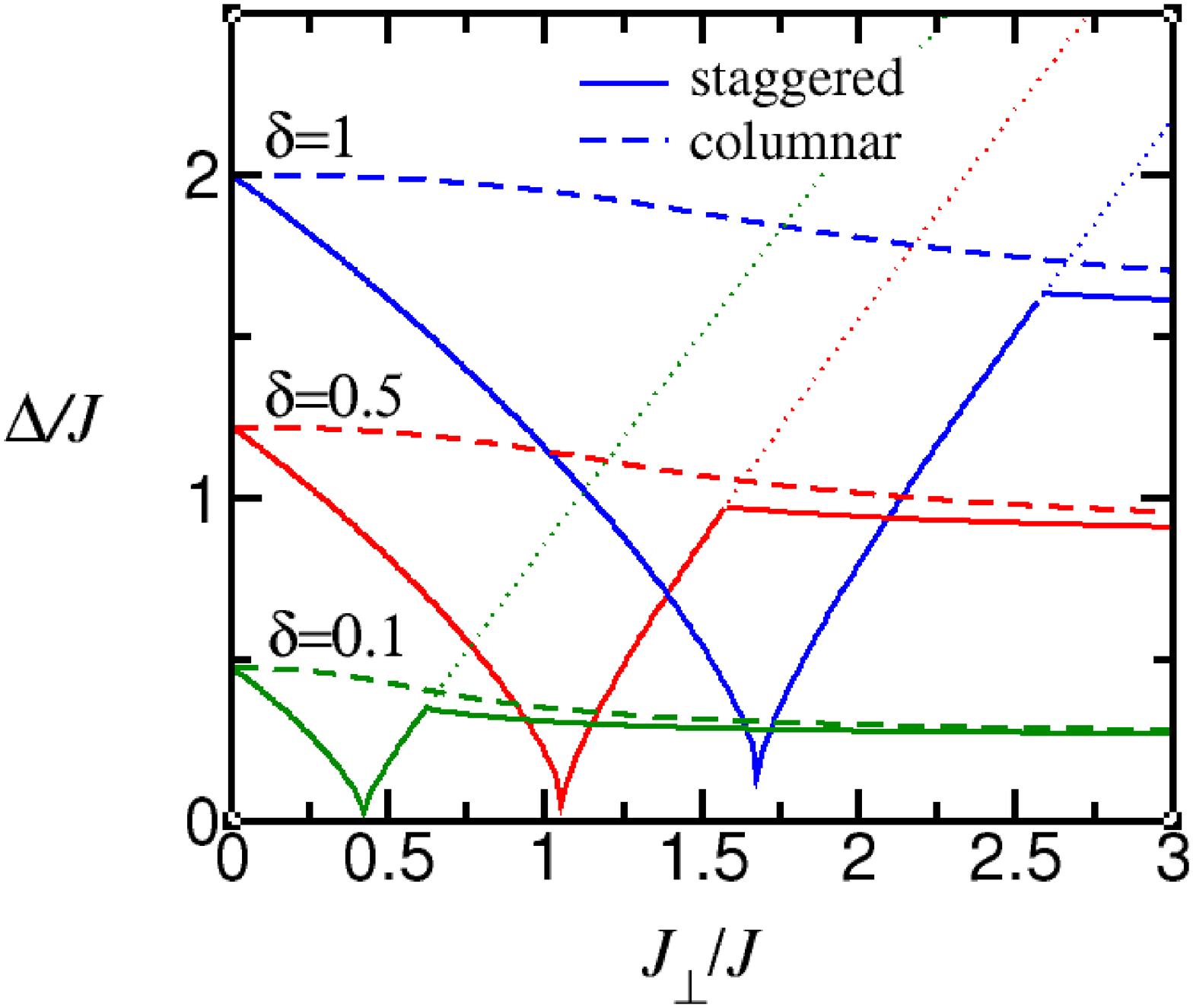,width=0.30\textwidth,angle=0} \caption{Three-leg
ladder: Gaps of the staggered and columnar phases.} \label{Gaps3L}
\end{figure}

The staggered phase is gapped everywhere,  \textit{except for a critical line
of vanishing gap, when $ \delta\neq 0$, $J_\bot \neq 0$.} Thus, we confirm the
original conjecture \cite{Delgado96}, which has been corroborated by the recent
DMRG calculations of Almeida \textit{et al} \cite{Almeida07b}. The gap of the
staggered phase, found from $E^{\rm s}_{3}(0)$, reads as
\begin{widetext}
\begin{equation}
 \label{Ds3Left}
  \Delta_s^< = \frac{J}{\sqrt{2}} \left[
   \delta^2(u^2+u'^2) + \Big( \frac{J_\bot p}{J} \Big)^2 -
   (u+u') \sqrt{(u-u')^2 \delta^2 + 2 \Big( \frac{J_\bot p}{J} \Big)^2 }
 \right]^\frac12,
\end{equation}
\end{widetext}
where $u'$ is given by Eq.~(\ref{s}) with replacement $Q_\pm \to Q_\pm'$.  The
columnar gap $\Delta_c$ and the (critical) staggered gap $\Delta_s$, directly
calculated from the mean-field equations, are plotted in Fig.~\ref{Gaps3L}.
%
%
There is one subtlety related to the staggered gap, absent in case of the
two-leg ladder. Recall that we defined gap in a given phase as a
\textit{minimal} band gap of the energy eigenvalues of the effective
single-particle Hamiltonian. What happens here is that the staggered gap
(\ref{Ds3Left}) is determined from the eigenvalue $E^{\rm s}_{3}$ which has
minimal band gap at small $J_\bot /J$. However with the increase of $J_\bot /J$
it reaches a certain value where ``a level crossing" occurs, i.e., the band
gaps of $E^{\rm s}_{3}$ and $E^{\rm s}_{1}$ are equal. Then, for bigger $J_\bot
/J$, the staggered gap is determined by the (minimal) band gap of $E^{\rm
s}_{1}$. In addition, since $E^{\rm s}_{1}= E^{\rm c}_{1}$ (\ref{E1}), the
staggered and columnar gaps are equal, and according to (\ref{Dc3}),
$\Delta_s^>= \Delta_c=J u \delta$ in that region. In fact, however, one can see
differences between $\Delta_s^>$ and $\Delta_c$. The reason is that the
self-consistently determined mean-field parameters entering Eq.~(\ref{s}) for
$u$, are not necessarily equal for two configurations.

\begin{figure}[h]
\epsfig{file=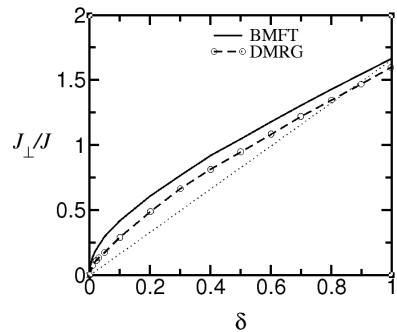,width=0.30\textwidth,angle=0} \caption{Three-leg
ladder: Critical line of the vanishing staggered gap. Our results are shown
along with the DMRG data from Ref.~\onlinecite{Almeida07b}.} \label{Crit3L}
\end{figure}

To get some analytically tractable formulas for the gap, let us approximate
$u=u'$. Then
\begin{equation}
\label{Ds3app}
    \Delta_s^< \approx J \big| u \delta - \frac{p}{\sqrt{2}} \frac{J_\bot }{J}  \big|
    ~.
\end{equation}
Making even a more drastic approximation for the bond mean-field parameters by
their maximal values, i.e., $u=2$ and $p=1+1/\sqrt{2}$ ($P=2^{-\frac32}$), we
get
\begin{equation}
\label{Ds3appM}
    \Delta_s^< \approx 2 J \big|  \delta - \frac{1+\sqrt{2}}{4} \frac{J_\bot }{J}  \big|
    ~.
\end{equation}
We plot in Fig.~\ref{Crit3L} the critical line obtained from the ``exact"
mean-field gap equation (\ref{Ds3Left}) and the DMRG result of
Ref.~\onlinecite{Almeida07b}. The dotted line  $J_\bot / J =1.6569 \delta$ is
the prediction of the simplified gap equation (\ref{Ds3appM}).
%
%
Note that contrary to the two-leg ladder, where we have exact predictions for
the two integrable end points of the critical line, for this case we don't have
an independent prediction for the critical ratio $J_\bot /J$ at
$\delta=1$.\cite{note2} We can point out however a very good agreement between
the mean field and DMRG. The former via Eq.~(\ref{Ds3Left}) yields 1.67 [or
1.66 via (\ref{Ds3appM})], while the latter yields 1.60.
%
%
\section{Conclusions}\label{Concl}
%
%
We analyze the possibility of quantum criticality (gaplessness) in dimerized
two- and three-leg ladders. Contrary to earlier studies of these models, we do
not imply a particular ladder's dimerization pattern. In this work we restrict
ourselves to the zero-temperature properties. We find that for a given
intrinsic intrachain dimerization, the ladder has lower zero-temperature energy
for the columnar dimerization order. This is true for both types of ladders
(two-, three-leg) considered. For the columnar dimerization pattern the ladders
are always gapped, i.e., there are no lines of quantum criticality on the
($\delta, J_\bot$) plane.

For the staggered dimerization order, we find that the ladders possess the
quantum critical lines, in agreement with earlier analyses of this problem.
\cite{Delgado98,Kotov99,Cabra99,Nersesyan00,Okamoto03,Almeida07a,Almeida07b}
The mean-field theory we apply in this study demonstrates its quantitative
accuracy. For the two-leg ladder, the mean-field critical line passes through
both integrable quantum critical points ($0,0$) and ($1,2 J$) on the ($\delta,
J_\bot$) plane, and demonstrates a good agreement with the numerical
diagonalization results \cite{Okamoto03} in between. For the case of three-leg
ladder, the mean-field critical line passes through ($0,0$) (the only
integrable quantum critical point) and agrees well with the critical line
determined through the DMRG calculations. \cite{Almeida07b}

Thus, from a more practical point of view, the very possibility of quantum
criticality in dimerized ladders hinges on some mechanism which would lock the
dimerization order into a more energetically expensive staggered configuration.
A realistic example of dimerization in a three-leg ladder through the
spin-Peierls mechanism results in the columnar order and so, the gapful phase
everywhere at non-zero dimerization. \cite{Khalada} At the moment we are not
aware of any Hamiltonian with would provide such a mechanism for quantum
criticality in dimerized ladders to occur.

%
%
\begin{acknowledgments}
We thank J. Almeida for useful correspondence. We acknowledge financial support
from the Natural Science and Engineering Research Council of Canada (NSERC) and
the Laurentian University Research Fund (LURF).
\end{acknowledgments}
%



\begin{thebibliography}{}
\bibitem{noteAF} In this work we study the case of the antiferromagnetic
(positive) spin exchange couplings only.
\bibitem{noteDH} Note that contrary to the sine-Gordon formula (\ref{ESG}), the
logarithmically corrected dimerized chain energy (\ref{Edhc}) is lower than the
plaquette energy (\ref{EplNU}) in the region $\delta_{\mathrm{eff}} \gtrsim
0.4$. However, the formulas (\ref{Edhc},\ref{EplNU}) are reliable only when
$\delta_{\mathrm{eff}} \ll 1$.
\bibitem{noteG}
The other band $E^{\rm s}_2$ can be excluded from further
considerations, since it has a bigger non-vanishing gap.
\bibitem{note2}
Using the formula for the NL$\sigma$M topological term derived in
Ref.~\onlinecite{Delgado96}, we obtain the critical line located
between two points: ($\delta=\frac{18}{35} = 0.5142...$, $J_\bot
=0$) and ($\delta=1$, $J_\bot/J =1.9823...$). So, this approach does
not result in adequate predictions.
\end{thebibliography}
\end{document}